\documentclass[preprint,12pt]{elsarticle}

\usepackage{amssymb}
\usepackage{bm}
\usepackage{amsmath}

\journal{Physics Letters A}

\begin{document}

\begin{frontmatter}

\title{Super Bloch oscillations in the Peyrard-Bishop-Holstein model}

\author{C.\ Herrero-G\'{o}mez, 
E.\ D\'{\i}az\footnote{Corresponding author. Tel.: +34 91 394 4747; fax: +34 91 394 4547.
E-mail address: elenadg@fis.ucm.es (Elena D\'{\i}azD).} 
and F.\ Dom\'{\i}nguez-Adame}

\address{GISC, Departamento de F\'{\i}sica de Materiales, Universidad
Complutense, E-28040 Madrid, Spain}

\begin{abstract}

Recently, polarons in the Peyrard-Bishop-Holstein model under DC electric fields 
were established to perform Bloch oscillations, provided the charge-lattice coupling is not large. 
In this work, we study this model when the charge is subjected to an applied
field with both DC and AC components. Similarly to what happens in the
rigid lattice, we find that the carrier undergoes a directed motion or
coherent oscillations when the AC field is resonant or detuned with respect to
the Bloch frequency, respectively. The electric density current and its Fourier
spectrum are also studied to reveal the frequencies involved in the polaron dynamics.

\end{abstract}


\end{frontmatter}

\section{Introduction}
\label{intro_PBH_SBO}

In perfect crystals whose lattices generate a periodic potential, Bloch theorem
predicts uniformly states extended over the whole system~\cite{Harrison80}.
However,  if a uniform electrical field $F$ is applied, in addition to this
periodic potential, all states become spatially localized due to Bragg
reflections~\cite{Bloch28,Zener34,Dunlap88}. In such a case, neglecting 
scattering effects, electron keep oscillating within a finite volume. The
frequency $\omega_B$ and the amplitude $L_B$ of the oscillation can be
established from semiclassical arguments~\cite{Esaki70,Ashcroft76}. The former
is usually known as Bloch frequency and it is proportional to the applied
electric field $F$, namely $\omega_{B}=eFa/\hbar$, where $-e$ is the electron
charge and $a$ refers to the lattice period along the field direction.  This
periodic motion takes place in real and in $k$ space, and it is known as Bloch
oscillation (BO). Since scattering processes, due to phonons or defects for
instance, destroy the coherence necessary to support BOs, its experimental
detection is highly  nontrivial. The first BOs were detected in semiconducting
superlattices~\cite{feld,wasch,deko1,martini,los}. More recently, cold atoms and
Bose-Einstein condensates (BECs) in optical lattices have been revealed as a
very convenient scenario to observe BOs since scattering processes can be
significantly reduced~\cite{dah,wilk,anders}. However, decoherence effects
cannot be removed completely and should be taken into account from a
theoretical  point of view. For instance, in BECs the atom-atom interaction
gives rise to well-known dynamical instabilities which can destroy the coherence
required to observe stable BOs~\cite{Roati04,Fallani04,Trombettoni01}. Still,
there have been some experimental and theoretical proposals to avoid these
undesirable effects~\cite{Gustavsson08,Gaul09}. Also, in organic molecules,
which are very flexible, the vibrations of the lattice are relevant and could
rapidly degrade the electron quantum coherence. In this regard, the
Peyrard-Bishop-Holstein (PBH) model of charge transport in DNA~\cite{Maniadis05}
was considered to demonstrate that polarons perform BOs even at realistic values
of the carrier-lattice coupling~\cite{Diaz08}.

In the last decade, a new interest in the dynamics of a quasiparticle affected
not only by a DC field but also by a superimposed AC field has 
emerged~\cite{Kolovsky10,Thommen02,Alberti09,Ivanov08}. Remarkably, it was
demonstrated experimentally that a weakly interacting BEC in a harmonically
driven tilted potential can support directed transport or large amplitude
oscillations of the  wave function, depending on the driven
frequency~\cite{Haller10}. The latter are known as super Bloch oscillations
(SBOs) and their characterizing parameters can also be described
semiclassically~\cite{Caetano11}. In this work we use the PBH model to study how
the introduction of the carrier-lattice interaction affects the dynamics of
polarons in this scenario. We will also calculate the current density associated
to the charge motion occurring under the applied fields and its Fourier spectrum
to reveal the main frequencies involved in the polaron dynamics. 

\section{Model}
\label{model_PBH_DNA}

The Hamiltonian of the PBH model can be written as
\begin{equation}
{\cal H}= {\cal H}_\mathrm{lat}+{\cal H}_\mathrm{car}+{\cal H}_\mathrm{int}\,.
\label{ham_PBH_DNA}
\end{equation}
The first term describes a one-dimensional anharmonic lattice according to the
Peyrard-Bishop model~\cite{Peyrard89}. A single degree of freedom $x_n$ is
assigned to every site, taking into account its displacement from the
equilibrium configuration. ${\cal H}_\mathrm{lat}$ for a homogeneous lattice
reads
\begin{equation}
{\cal H}_\mathrm{lat}=\sum_{n}^{N}
\Big[\frac{1}{2}\,m \dot{x}^2_n + V(x_n)+W(x_n,x_{n-1}) \Big ]\, ,
\label{Hlat}
\end{equation}
where $m$ is the mass of each site and $n=1,2,\ldots N$ labels the sites along
the system. Two potential terms appear in this Hamiltonian, namely a local
Morse potential $V(x_n)$ and a nonlinear anharmonic coupling  between
nearest-neighbors $W(x_n,x_{n-1})$
\begin{subequations}
\begin{equation}
V_M(x_n)=V_0\left(e^{-\alpha x_{n}}-1\right)^2,
\label{Morse}
\end{equation}
\begin{equation}
W(x_n,x_{n-1})=\frac{k}{4}\big(2+e^{-\beta(x_{n}+x_{n-1})}\big)
\big(x_{n}-x_{n-1}\big)^2\ .
\label{W}
\end{equation}
\end{subequations}
For the sake of concreteness, the fitting parameters are chosen according to 
those obtained to reproduce experimental DNA
melting curves within the Peyrard-Bishop model, $m=300\,$amu, $V_0=0.04\,$eV,
$\alpha=4.45\,$\AA$^{-1}$, $k=0.04\,$eV/\AA$^{2}$ and
$\beta=0.35\,$\AA$^{-1}$~\cite{Dauxois93}.

The charge carrier Hamiltonian ${\cal H}_\mathrm{car}$ in the unbiased  lattice
is expressed within the nearest-neighbor approximation as
follows~\cite{Komineas02}
\begin{equation}
{\cal H}_\mathrm{car}= -T\sum_{n}^{N}  
\left(a_{n}^{\dag} a_{n+1}^{}+a_{n+1}^{\dag} a_{n}^{}\right)\, ,
\label{Hch}
\end{equation}
where $T$ is the nearest-neighbor hopping and $a_{n}^{\dag}$ and $a_{n}$ 
denotes the carrier creation and annihilation operators, respectively. Since
there is not a  common value for the parameter $T$ in the literature, we will
take $T=0.1\,$eV hereafter as a good representative
one~\cite{Yan02,Voityuk02,Senthilkumar05}.

The last term in~(\ref{ham_PBH_DNA}) takes into account a Holstein-like
carrier-lattice interaction as an on-site energy correction as
follows~\cite{Komineas02}
\begin{equation}
{\cal H}_\mathrm{int}= -\chi\sum_{n}^{N} x_n^{} a_{n}^{\dag} a_{n}^{}\,.
\label{Hint}
\end{equation}
Here $\chi$ denotes the carrier-lattice coupling constant. \emph{Ab-initio\/} 
estimations of this coupling $\chi$ are scarce and therefore we will vary its
magnitude in our numerical simulations.

In typical molecular systems it is possible to use a semiclassical approach due
to the different time-scales of the  charge and the lattice
dynamics~\cite{Kalosakas98}. Thus, the dynamics of the carrier under a field
with both DC and AC components such as  $F(t)=F_0+F_1 \sin(\omega t +\phi)$ can
be studied by way of the following Schr\"{o}dinger equation~\cite{Maniadis05}
\begin{equation}
i\hbar\,\frac{d\psi_n}{dt}=-\big[U_0+U_1 \sin(\omega t +\phi)\big]n
\psi_n-T(\psi_{n+1}+\psi_{n-1})+\chi x_n\psi_n \ ,
\label{Schrodinger}
\end{equation}
where $\psi_n$ is the probability amplitude for the charge carrier located at the
$n$th site. The parameters $U_{i}=eaF_{i}$ with $i=0,1$ are the energy terms
associated  to the applied electric fields  $F_{0,1}$, $a$ being the period of
the lattice (e.g. $a=3.4\,$\AA\ in DNA). The last term in
Eq.~(\ref{Schrodinger}) describes the carrier-lattice coupling through the
constant $\chi$ and the displacement $x_n$ from its equilibrium position.
Newton's equations of motion for the displacements $x_n$ become
\begin{equation}
m\,\frac{d^2x_n}{dt^2}=-V_{M}^{\prime}(x_n)-W^{\prime}(x_n,x_{n-1})
-W^{\prime}(x_{n},x_{n+1})-\chi |\psi_n|^2 \ ,
\label{Newton}
\end{equation}
where the prime indicates differentiation with respect to $x_n$.
In what follows we will take the stationary polaron of the unbiased system under
consideration~\cite{Komineas02} as the initial condition for the integration of
Eqs.~(\ref{Schrodinger}) and~(\ref{Newton}) in a lattice subjected to the 
electric field $F(t)$. 

\section{Motion of the polaron in a biased system}
\label{motion_PBH_SBO}

In Ref.~\cite{Haller10} the dynamics of a BEC under superposed DC and AC forces
was observed experimentally. Later, the authors of Ref.~\cite{Caetano11} studied
the wave packet dynamics under these conditions in a rigid lattice ($\chi=0$) by
means of a semiclassical approach and numerical calculations. In both works it
was reported that for a resonant AC field with frequency $\omega=\omega_B$,  the
wave function of the carrier performs a directed motion with oscillatory
features with drift velocity depending on the phase $\phi$. On the contrary, if
the frequency of the AC field is detuned with respect to $\omega_B$, the
wave packet exhibits a beating effect called SBOs~\cite{Haller10}. These two
situations will be referred to as resonant and detuned cases hereafter. In this
work, we will focus on two set of parameters as representative values of these
two different situations, for several values of the carrier-lattice coupling. 

\begin{figure}[h]
\centerline{
\begin{minipage}{.40\linewidth}
          \includegraphics[width=50mm,clip]{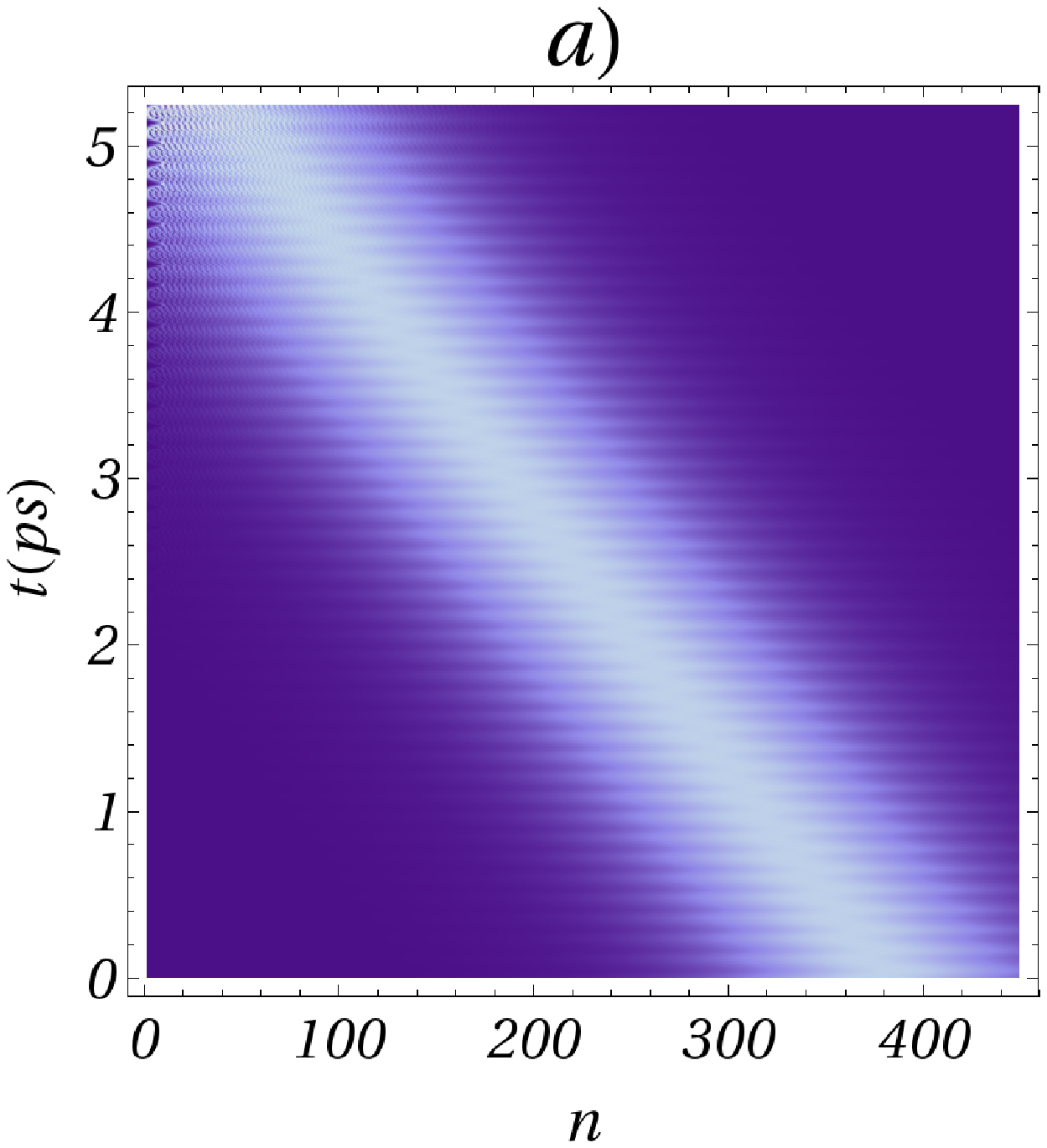}
    \end{minipage}%
\begin{minipage}{.40\linewidth}
          \includegraphics[width=50mm,clip]{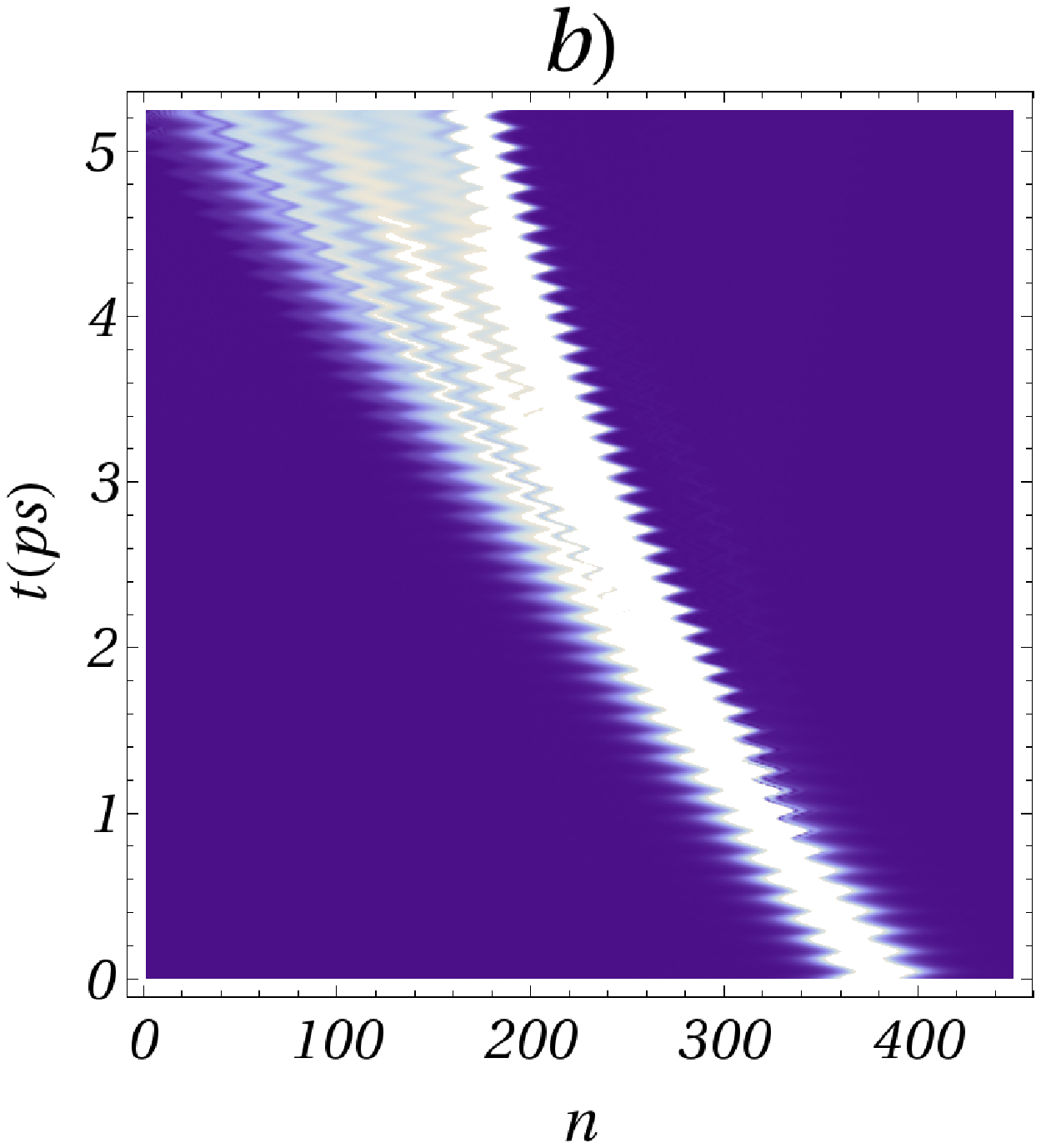}
    \end{minipage}
}
\centerline{\begin{minipage}{.40\linewidth}
          \includegraphics[width=50mm,clip]{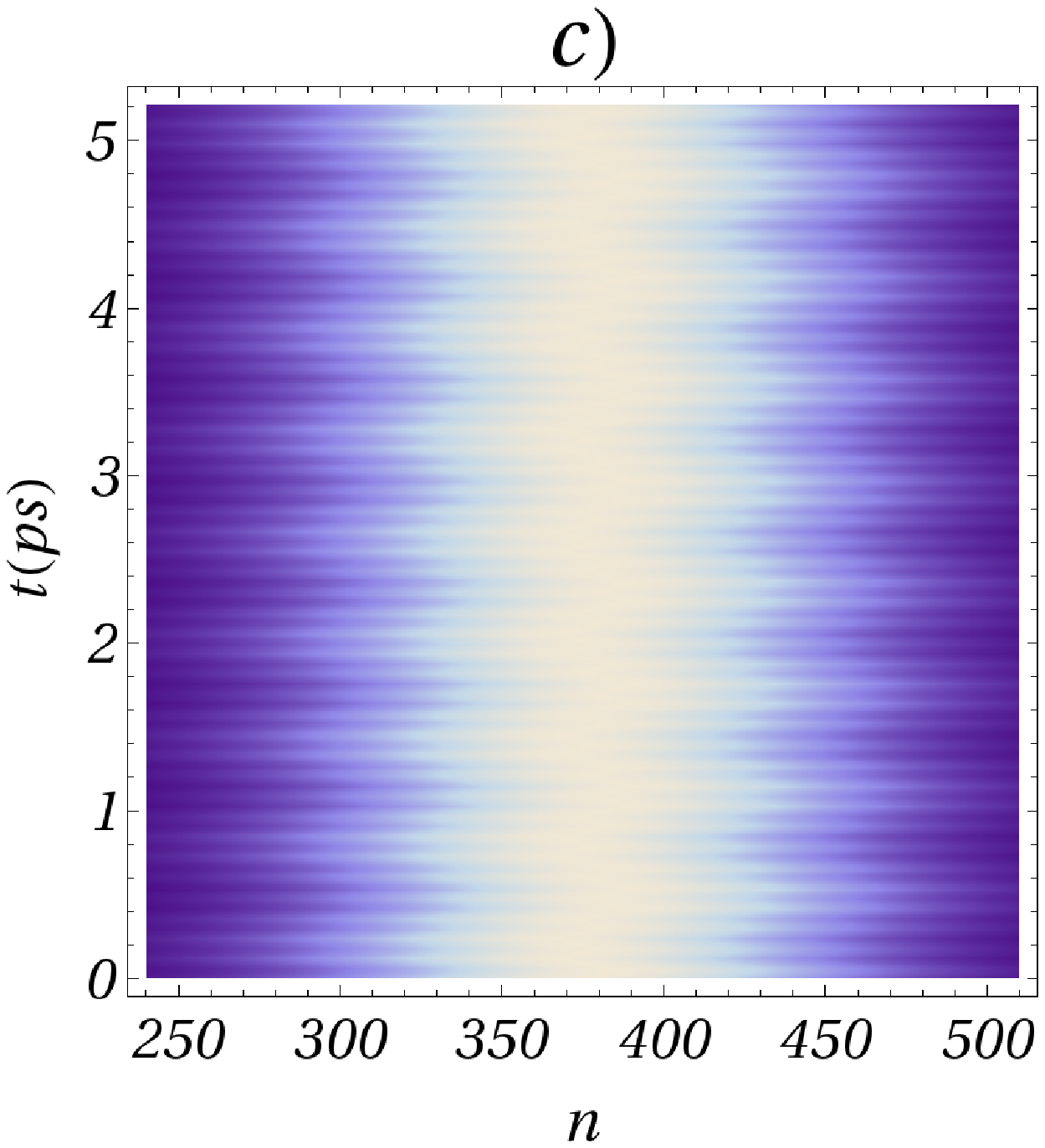}
    \end{minipage}%
\begin{minipage}{.40\linewidth}
          \includegraphics[width=50mm,clip]{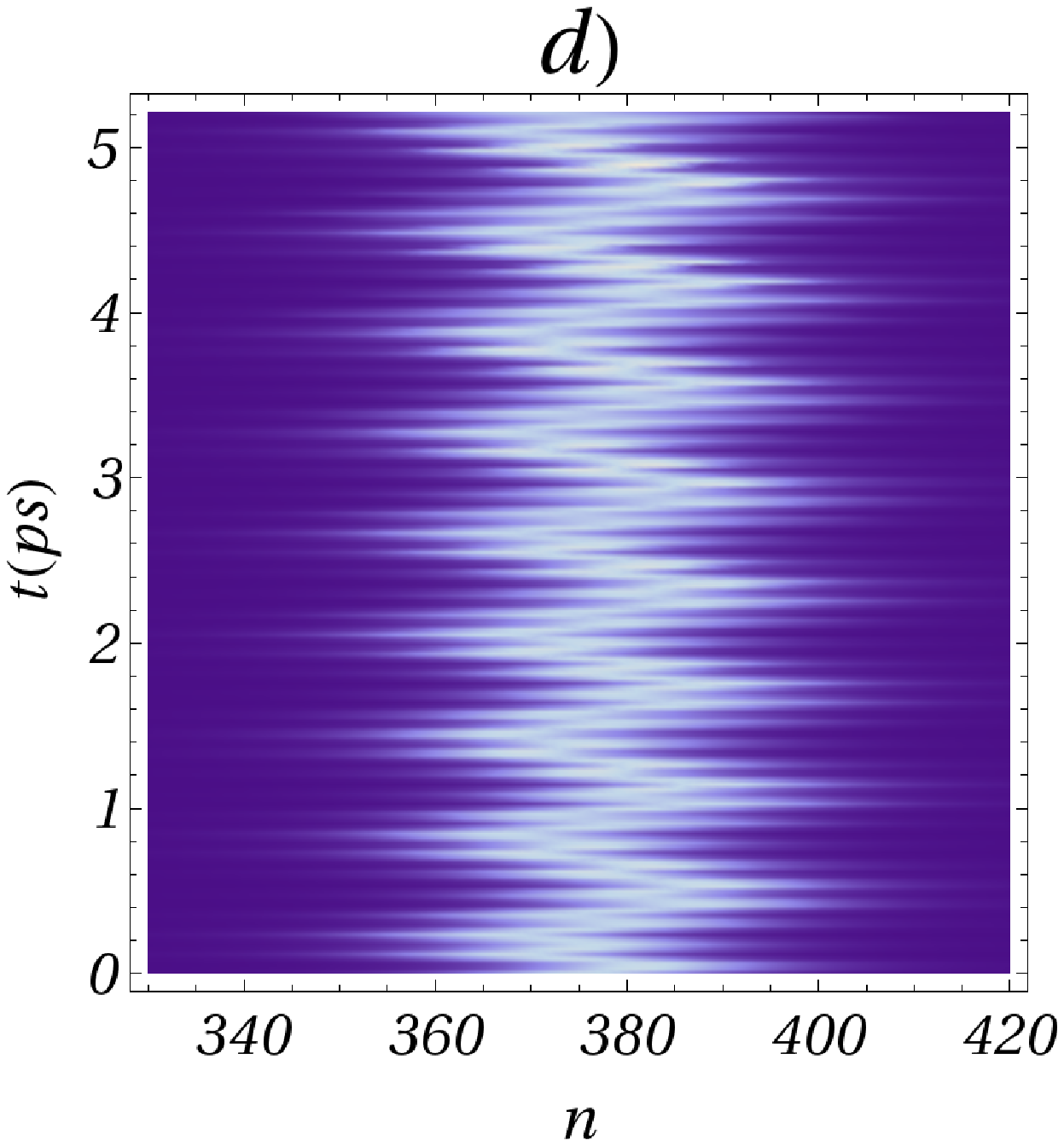}
    \end{minipage}
}
\caption{Modulus of the carrier wave function in a lattice of $N=750$ sites as a
function of position and time with $F_0=10\,$meV/\AA\, $F_1=0.5 F_0$ and
$\chi=0.1\,$eV/\AA\ (left panels) and $\chi=0.3\,$eV/\AA\ (right panels).  The
two upper panels correspond to resonant cases with $\omega=\omega_B$ and
$\phi=0$. The two lower panels correspond to detuned cases with $\omega=1.2
\omega_B$ and $\phi=0$. Light and dark regions indicate nonzero and zero values,
respectively.}
\label{Fig1} 
\end{figure}

First, we analyze how the polaron evolves when the field
$F(t)=F_0+F_1\sin(\omega t +\phi)$ is applied. The time-domain evolution of the
carrier wave packet obtained by direct integration of Eqs.~(\ref{Schrodinger})
and~(\ref{Newton}) is shown in Fig.~\ref{Fig1} for the resonant and the detuned
cases, and $\chi=0.1$ and $0.3\,$eV/\AA. 

Similarly to the case of the rigid lattice, the two upper panels  of
Fig.~\ref{Fig1} (resonant cases with $\omega=\omega_B$) show the directed motion
of the polaron as well as superimposed BOs. On the contrary, the two lower
panels of Fig.~\ref{Fig1} (detuned cases with $\omega=1.2\omega_B$) shows that
the carrier perform SBOs. It is to be noticed that by increasing the strength of
the carrier-lattice coupling the localization of the stationary states becomes
larger and therefore the wave packet motion is more clearly observed (see
Fig.~\ref{Fig1}d). We would like to stress that contrary to what happens in the
rigid lattice~\cite{Caetano11}, due to the nonlinear interactions the wave
packet remains localized in time except for long time-scales if the
carrier-lattice coupling is large, as seen in Fig.~\ref{Fig1}b). 

The dynamics of the polaron can be monitored in more detail by means of the
centroid of the carrier wave function $c(t)=\bar{n} (t)-\bar{n}(0)$ with
$\bar{n}(t)=\sum_{n} n |\psi_{n}(t)|^2$. Similarly we define the dimensionless
magnitude $l(t)=\xi(t)-\xi(0)$ with $\xi(t)=\sum_{n=1}^{N} n\,x_{n}(t)/a$ for
the lattice displacements. We also study the Fourier transform of these
magnitudes to reveal the frequencies involved in the dynamics of the polaron.

\begin{figure}[h]
\centerline{\includegraphics[width=100mm,clip]{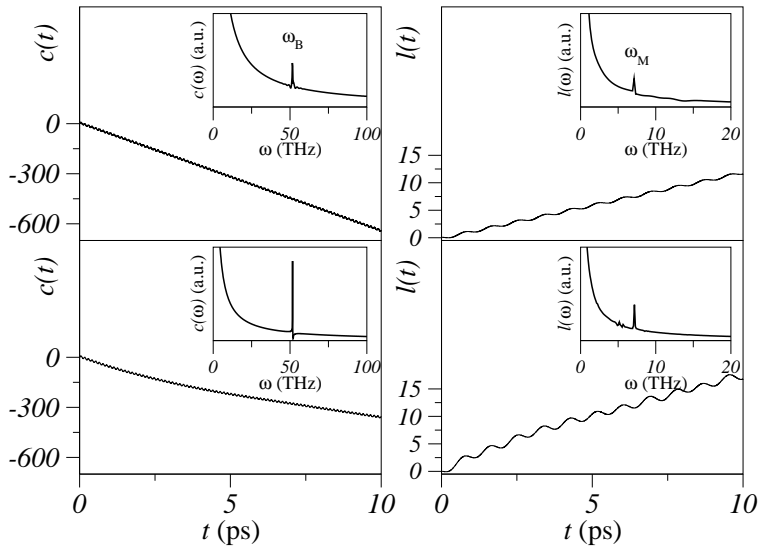}} \caption{Centroid of
the carrier wave function (left) and $l(t)$ (right) as a function of time in a
lattice of $N=2000$ sites for $\omega=\omega_B$, $\phi=0$ and $\chi=0.1\,$eV/\AA\ (upper
panel) and $\chi=0.3\,$eV/\AA\ (lower panel). Insets show the corresponding
Fourier spectra.}
\label{Fig2} 
\end{figure}

Figure~\ref{Fig2} shows that $c(t)$ and $l(t)$ perform a directed motion with 
short-time-scale oscillations in a lattice of $N=2000$ sites. The Fourier
spectrum $c(\omega)$ indicates that the main frequency of such oscillations is
equal to  $\omega_B= eaF_0/\hbar = 51.67\,$THz (very weak peaks at multiples of $\omega_B$ are also found),
while for $l(\omega)$ the Morse
frequency $\omega_M=\sqrt{2V_0/m}=7.22\,$TH is the relevant one. Notice that the
$\omega_M$ is the harmonic frequency of the small amplitude oscillations of a
mass $m$ around the minimum of the Morse potential~(\ref{Morse}). 

\begin{figure}[h]
\centerline{\includegraphics[width=100mm,clip]{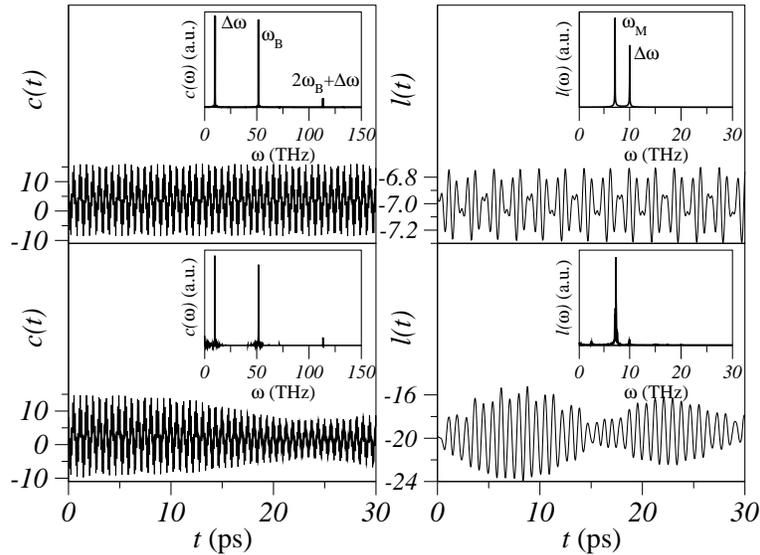}}
\caption{Same as in Fig.~\ref{Fig2} but for detuned frequency
($\omega=1.2\omega_B$ and $\phi=0$) in a lattice of $N=750$.}
\label{Fig3} 
\end{figure}

The results corresponding to the detuned cases ($\omega=1.2\omega_B$) are shown
in Fig.~\ref{Fig3}. $c(t)$ and $l(t)$ display a more complex dynamics and
absence of directed motion. In the case of $c(t)$ the dynamics corresponds to
the SBOs with a large period $2\pi/\Delta\omega$ defined by the detuning
frequency $\Delta\omega=0.2\omega_B$ and a short period $2\pi/\omega_B$. In
addition, in  $c(\omega)$ we obtained other peaks which can be described with
the full analytical solution of the centroid motion in the rigid
lattice~\cite{Kolovsky10} [see the insets of the left panels of
Fig.~\ref{Fig3}]. The main frequency involved in $l(t)$ is again the Morse
frequency but now we also observe the occurrence of a smaller peak at the
detuning frequency [see the insets of the right panels of Fig.~\ref{Fig3}].
Therefore, we come to the conclusion that in both situations the oscillations of
the lattice and the carrier are almost decoupled, at least at moderate applied
fields.

\section{Average current density}
\label{current_PBH_SBO}

In view of the oscillating behavior of the carrier wave packet described in the
previous section, it seems reasonable to expect that the electric current 
behaves in a similar way. Therefore, and since the current is a macroscopic
magnitude which can be directly observed in experiments, we calculate the
average current density $J(t)$ according to the following
expression~\cite{Maniadis05}
\begin{equation}
J(t)=\frac{\hbar e}{m_cNa^2}\sum_{n=1}^N 
\mathrm{Im}\big[\psi_n^*(\psi_{n+1}-\psi_{n-1})\big]\ ,
\label{J_PBH_DNA}
\end{equation}
where $m_c$ is the mass of the carrier.

\begin{figure}[h]
\centerline{\includegraphics[width=100mm,clip]{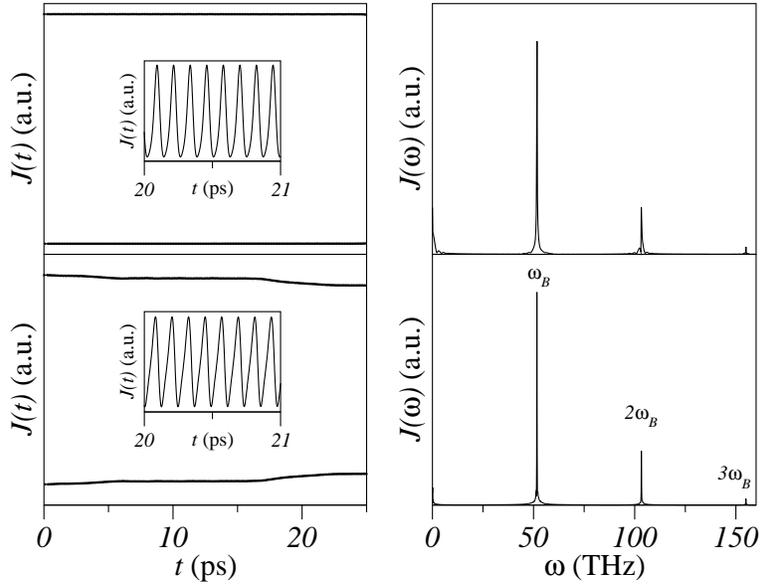}}
\caption{Left panels show the envelope of the average current density as a
function of time in a lattice of $N=2000$ sites for the resonant case for
$\omega=\omega_B$, $\phi=0$ and $\chi=0.1\,$eV/\AA\ (upper panel) and $\chi=0.3\,$eV/\AA\
(lower panel). Insets show the short time behavior of the average current
density. The right panels show the corresponding Fourier transform of the
average current density.}
\label{Fig4} 
\end{figure}

\begin{figure}[ht]
\centerline{\includegraphics[width=100mm,clip]{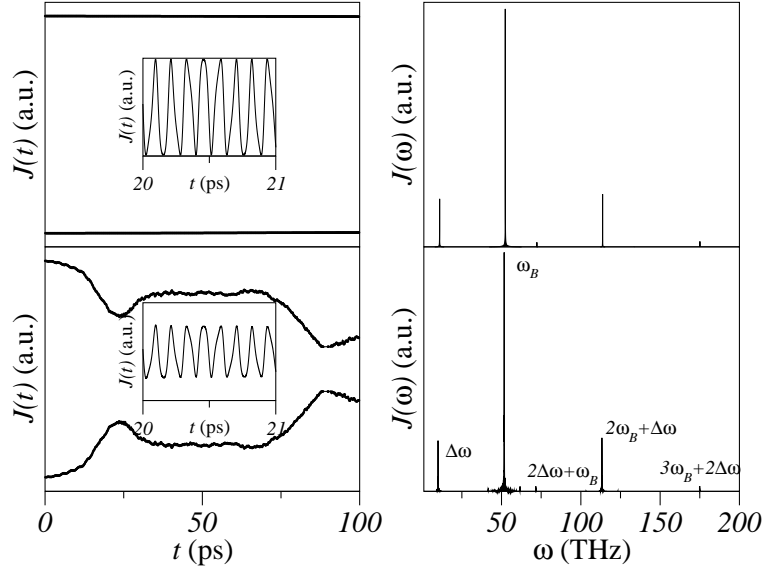}}
\caption{Same as in Fig.~\ref{Fig4} but for detuned frequency
($\omega=1.2\omega_B$ and $\phi=0$) in a lattice of $N=750$.}
\label{Fig5} 
\end{figure}

As in the previous section, we will consider different values of the
carrier-lattice coupling for the representative resonant and detuned case. 
Typical results of our simulations are collected in Figs.~\ref{Fig4}
and~\ref{Fig5}. The left panels show the envelope of the  average current
density~(\ref{J_PBH_DNA}) over a long time interval while the insets display the
short time behavior for two different values of the coupling constant 
$\chi=0.1\,$eV/\AA\ and $\chi=0.3\,$eV/\AA. 

The average current density displays a well-defined oscillatory behavior in both
cases, whose relevant frequencies match perfectly those obtained from the
centroid motion in Figs.~\ref{Fig2} and~\ref{Fig3}. Notice that these
frequencies are in agreement to those analytically predicted for the rigid
lattice in all cases~\cite{Kolovsky10,Caetano11}. We stress that the increase of
the carrier-lattice coupling leads to a faster modulation of the average current
density but, remarkably, the oscillations do not decay in time. 

\section{Conclusions}   
\label{conclusions_PBH_SBO}

We have studied the dynamics of the carrier dynamics in the PBH model under
superimposed DC and AC fields and found that it is similar to that found in the
rigid lattice~\cite{Haller10,Caetano11}. The carrier-lattice coupling leads to a
distortion of the initial shape of the wave packet at long times and to a
faster  modulation of the centroid motion. Still, the polaron display SBOs which
do not decay in time when the driven frequency is detuned with respect to the
Bloch frequency.  The carrier is partially decoupled from the lattice, whose
relevant frequency is not the Bloch frequency but the one associated to the
Morse potential.

\section*{Acknowledgments}

This work was supported by MICINN (projects MAT2010-17180 and MOSAICO). C.~H.
acknowledges financial support by MEC (Program Becas de Colaboraci\'{o}n).

\end{document}